\documentclass{aa}
\usepackage{graphicx}
\usepackage{graphicx}
\usepackage{bm,natbib,url,wasysym}
\usepackage[varg]{txfonts}
\bibpunct{(}{)}{;}{a}{}{,} 

\graphicspath{{./fig/}{./png/}}

\usepackage{color}
\usepackage[normalem]{ulem}

\newcommand{\EQ}{\begin{equation}}
\newcommand{\EN}{\end{equation}}
\newcommand{\EQA}{\begin{eqnarray}}
\newcommand{\ENA}{\end{eqnarray}}

\newcommand{\Eq}[1]{Equation~(\ref{#1})}
\newcommand{\Eqs}[2]{Equations~(\ref{#1}) and~(\ref{#2})}
\newcommand{\Eqss}[2]{Equations~(\ref{#1})--(\ref{#2})}

\newcommand{\Sec}[1]{Section~\ref{#1}}

\newcommand{\Fig}[1]{Fig.~\ref{#1}}

\newcommand{\Figs}[2]{Figs.~\ref{#1} and \ref{#2}}

{}
{}
{}

{}
{}
{}
{}
{}
{}
{}
{}
{}
{}
{}
{}
{}
{}
{}
{}
{}
{}

{}
{}
{}

{}

{}
{}

\newcommand{\uu}{\mbox{\boldmath $u$} {}}

\newcommand{\BB}{\mbox{\boldmath $B$} {}}

\newcommand{\JJ}{\mbox{\boldmath $J$} {}}

\newcommand{\nab}{\mbox{\boldmath $\nabla$} {}}

\def\BBind{\BB^{\rm ind}}
\def\Bind{B^{\rm ind}}

\newcommand{\Mm}{\,{\rm Mm}}

\begin{document}

\titlerunning{Current systems of coronal loops}
\authorrunning{Warnecke et al.}

\title{Current systems of coronal loops in 3D MHD simulations}
\author{J. Warnecke\inst{1} \and F. Chen\inst{2,1} \and
  S. Bingert\inst{3} \and H. Peter\inst{1}}
\institute{Max-Planck-Institut für Sonnensystemforschung,
  Justus-von-Liebig-Weg 3, 37077 G\"ottingen, Germany\\
\email{warnecke@mps.mpg.de}\label{inst1}
\and High Altitude Observatory, National Center for Atmospheric
Research, 3080 Center Green Dr., Boulder, CO 80301, USA\label{inst2}
\and  Gesellschaft für wissenschaftliche Datenverarbeitung mbH
  Göttingen, Am Fa\ss berg 11, 37077 G\"ottingen\label{inst3}}
\date{Received 18 November 2016 / Accepted 19 May 2017}
\abstract{}{%
We study the magnetic field and current structure
associated with a coronal loop. Through this we investigate to what
extent the assumptions of a force-free magnetic field break down and
where they might be justified.
}{%
We analyze a three-dimensional (3D) magnetohydrodynamic (MHD) model of
the solar corona in an emerging active region with the
focus on the structure of the forming coronal loops. The lower boundary of this
simulation is taken from a model of an emerging active region. As a
consequence of the emerging magnetic flux and the horizontal motions
at the surface a coronal loop forms self-consistently.
We investigate the current density along magnetic field lines inside (and outside)
this loop and study the magnetic and plasma properties in and around
this loop. The loop is defined as the bundle of field lines that
coincides with enhanced emission in extreme UV.
}{%
We find that the total current along the emerging loop changes its sign from
being antiparallel to parallel to the magnetic field.
This is caused by the inclination of the loop together with the
footpoint motion.
Around the loop, the currents form a complex non-force-free helical
structure.
This is directly related to a bipolar current structure at the loop
footpoints at the base of the corona and a local reduction of the
background magnetic field (i.e., outside the loop) caused by the plasma
flow into and along the loop.
Furthermore, the locally reduced magnetic pressure in the loop allows
the loop to sustain a higher density, which is crucial for the
emission in extreme UV. The action of the flow on the magnetic field
hosting the loop turns out to also be responsible for the observed
squashing of the loop.
}{
The complex magnetic field and current system surrounding it can only be
modeled in 3D MHD models where the magnetic field
has to balance the plasma pressure.
A one-dimensional coronal loop model or a force-free extrapolation cannot capture the  current system and the complex interaction of the
plasma and the magnetic field in the coronal loop, despite the
fact that the loop is under low-$\beta$ conditions.
}
\keywords{Magnetohydrodynamics (MHD) -- Sun: corona -- Sun: magnetic fields
}

\maketitle

\section{Introduction}\label{S:intro}

The solar corona is characterized by its high temperature and low
plasma density.
Mostly, there the magnetic field dominates the structures and dynamics
of the coronal plasma as quantified by a low value of plasma-$\beta$.
This is the ratio of magnetic to gas pressure.
Under the assumption of a low-$\beta$ corona, the magnetic field can
be modeled using force-free extrapolation of the photospheric magnetic
field \citep[for a review, we refer to][]{Wi08}.
In these models, the force-freeness is ensured by requiring the
currents to be parallel or antiparallel to the magnetic field, which
turns out to be not always valid \citep{PWCC15}.

Since the coronal model of \cite{GN02,GN05a} three-dimensional (3D)
magnetohydrodynamics (MHD) models can produce a loop-dominated corona
in a realistic setup.
In particular  these models generate self-consistently high coronal
temperatures through the braiding of magnetic field lines
\citep{P72} driven by photospheric motions of the magnetic footpoints
\citep[e.g.,][]{BP11}.
Such models produce a distribution of energy input
consistent with the nanoflare model \citep{BP13}.
The coronal emission synthesized from these models fits well with
observed coronal loop properties.
This applies to the spectroscopic properties, for example, average Doppler
shifts \citep{PGN04,PGN06,HHDC10}, the distribution of the emission
within an active region or along a loop \citep[e.g.,][]{MMLL05,MMLL08},
or the appearance of (small) loops based on the observed photospheric
magnetic field \citep{BBP13}.
Therefore these models can be considered as a good representation of
coronal structures and dynamics on the resolved scale \citep[e.g.,][]{P15}.
The magnetic field in these simulations turns out to be close to
force-free. This is indicated  by the exponential drop of both the
averaged heating rate (${\propto}j^2$) and magnetic energy
(${\propto}B^2$) with the same scale height \citep{GN05a}
(here $j$ and $B$ are the current density and the magnetic field
strength, respectively).
More direct support for the force-freeness is through the small angle
between the current and the magnetic field, which is on average smaller
than 20$^\circ$ in a stable active region in the model of
\cite{BP11}, shown in \cite{Bingert2009}.
However, a force-free magnetic field can also be achieved by
a simplified high plasma-$\beta$ coronal model \citep{WB10}.

\begin{figure}[t!]
\begin{center}
\includegraphics[width=0.9\columnwidth]{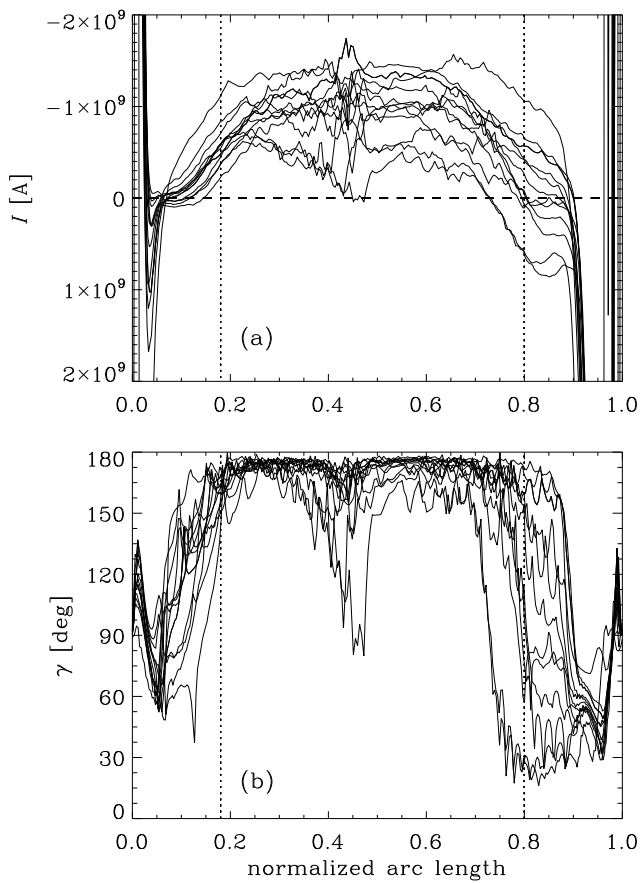}
\end{center}\caption[]{Testing the force-freeness in a 3D MHD model of
  a stable active region.
Panel (a) shows the total current, $I$, along the magnetic field,
$\BB$, where $I$ is integrated over the cross-section of the expanding
flux tube. Panel (b) displays the angle $\gamma$ between the current
density, $\JJ$, and $\BB$. Both $I$ and $\gamma$ are plotted for
twelve selected field lines in an EUV loop over the normalized arc
length (i.e., 0 and 1 are in the photosphere and the apex is at about
0.5).
$I$ and $\gamma$ are defined in \Eqs{eq:I}{eq:gam},
respectively.
The hot coronal part of the loop (with temperatures above 1\,MK) is
found between the vertical dotted lines. Based on data from
\cite{BP11}. See \Sec{S:intro}.
}
\label{sven_totcur}
\end{figure}

The most crucial test to ensure the field is force-free is to check the angle
of the magnetic field to the electric currents, as well as the total current
along the loop.
While the angle should be close to zero or 180$^\circ$, of course,
furthermore the absolute value of the current should be approximately
constant.
Because the magnetic field is expanding with height, this latter
criterium implies
that the total current through the cross section of a flux tube
hosting the loop defined by a set
of field lines has to be constant along the loop. 
However, the crucial part is that the current should not change
sign.
We analyze this
briefly for the model of a stable
active region from \cite{BP11} in which a quasi-steady loop
forms and exists for about 30 minutes, similar to an observed loop.
We display the total current and the angle in \Fig{sven_totcur} for
twelve selected field lines
within the loops seen in extreme UV emission (EUV). From
\Fig{sven_totcur}a it is clear that indeed the
total current is more or less constant in the coronal part. Of course,
this is no longer the case in the
lower chromospheric and photospheric part of the loop.
Furthermore, the sign of the current, and therefore its direction, is mostly
positive in the coronal region.
This behavior is underlined by the distribution of the angle
$\gamma$ between the magnetic field and the current density, as shown in 
\Fig{sven_totcur}(b).
In the coronal region, $\gamma$ is preferential close to $180^\circ$,
indicating an antiparallel current along most of the magnetic field lines.
From this we conclude that the magnetic field in the hot dense loops of this
model is indeed force-free.
The (rough) alignment of the currents with the magnetic field is also
found in models with a simpler magnetic geometry, for example, in the MHD
model of \cite{GN96} of a loop in a straight geometry. In their
reduced MHD model \cite{RVED08} investigated what role MHD turbulence
plays in such a setup for the braiding of the magnetic field.

The question arises of whether or not a more dynamic model of an active region with a
changing magnetic field in the corona would still host a magnetic
field in a nearly force-free state. This motivates the analysis of a
3D MHD model of an emerging active region, and for this purpose we
employ the model by \cite{CPBC14,CPBC15}, who simulated the coronal
response to the formation of an active region through flux emergence
in the photosphere.
There, a coronal magnetic loop structure  forms resulting in a bright
loop seen in EUV at coronal temperatures.
While this is a model of an emerging region, a real emerging active
region on the Sun will be much more violent, and thus our
considerations for violations of the force-free state should be
considered as a lower limit. Where we see significant
deviation from a force-free magnetic field in this model, for the real Sun, much more
severe violations should be expected.

A further motivation for our study are the recently performed plasma laboratory experiments of
a coronal loop eruption \citep{OMYJKX11,MYJYFJSD15}.
In their setup, a hot plasma loop, considered to be the equivalent of a
coronal loop, is generated by a constant current through a plasma
 confined by a  complex loop-like magnetic field structure.
This leads to a strong Lorentz-force, which is partly balanced by their
plasma and in consequence causes an eruption.
In their setup the magnetic field turns out to be mostly non-force-free.
Even though these experiments are still far from the configuration of
the solar corona they provide an interesting path towards understanding
the eruption mechanism.
Our study might provide some new insight into the current systems and
magnetic field  in and around coronal loops and therefore
contribute to the improvement of the interpretation and further development of the experiments.

While the laboratory aspect has its own appeal, the main interest of
our study is to test to what extent a changing magnetic field in an
emerging active region is in a force-free state.
For this we investigate data from an existing 3D MHD model
(\Sec{S:model.data}), investigate the current systems around the loop
forming in this model (\Sec{S:elect.current}) and discuss its
connection with the magnetic field structure (\Sec{S:consequences}).
Then we show how the plasma properties can influence the magnetic
field structure and evolution (\Sec{S:flows.mag} and
\ref{S:squashing}).
This leads us to the limitations of the force-free assumptions for
emerging active regions in \Sec{S:conclusions}.

\section{Three-dimensional magnetohydrodynamics model of the active region corona}
\label{S:model.data}

\begin{figure}[t!]
\begin{center}
\includegraphics[width=\columnwidth]{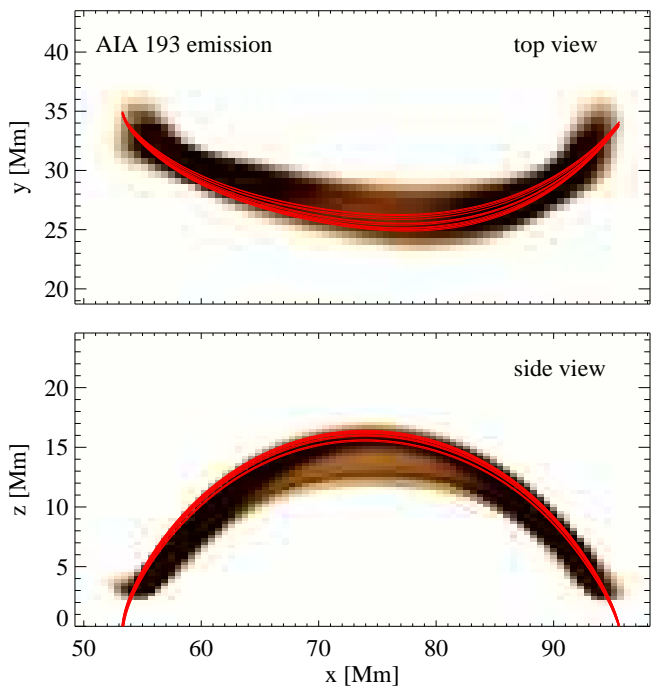}
\end{center}\caption[]{
Snapshot of coronal loop and field lines studied here. The background
image shows the synthesized EUV emission as it would be seen in the
193 \AA{} channel of AIA showing mainly 1.5\,MK hot
plasma. Overplotted are twelve selected  magnetic field lines (red)
within the EUV loop. This shows a snapshot at time  $t=12.5\,\min$
based on the data from the model by \cite{CPBC14,CPBC15}. The two
panels show a view from straight above (top panel) and the side
(bottom panel)  of part of the computational
domain; we refer to \Sec{S:model.data}. 
}
\label{em}
\end{figure}
In this work, we analyze data from the model described in detail in
\cite{CPBC14,CPBC15}.
The modeling strategy of that model is based on \cite{BP11,BP13}, but
all data shown here are from \cite{CPBC14,CPBC15}.
In short, in these 3D simulations the corona is modeled by solving the
equations of compressible  MHD with the {\sc Pencil
  Code}\footnote{{\tt http://github.com/pencil-code}} including the
induction equation, continuity equation, momentum equation, and energy
equation.
\cite{BP11,BP13} used the photospheric magnetogram from  an observed
well evolved active region as the lower boundary of the magnetic field
that is driven by horizontal motions.
This leads to a more or less stable active region corona.
In contrast, \cite{CPBC14,CPBC15} use the significantly changing
magnetic field (and velocity field)
in the photosphere from the flux emergence simulation of \cite{RC14}.
This flux emergence simulation is less violent than the emergence
model by \cite{Cheung_etal10} but still features the transition from
basically no magnetic field at the photosphere (our lower boundary) to
two strong opposite polarity sunspots. In the coronal model of
\cite{CPBC14,CPBC15} coupled to the flux emergence simulation this
results in the formation of a coronal loop structure that is
non-steady in nature.
Thus this coronal simulation is well suited to studying the currents
hosted in an active-region corona during phases of significant evolution.

We focus our study mainly on a snapshot at $t=12.5\,\min$
 of the coronal simulation of \cite{CPBC14,CPBC15}.\footnote{Actually this is more than an
   hour from the start of the coronal simulation when the first clear
   loop has formed. For the definition of the zero-time we refer to
   \cite{CPBC14,CPBC15} and their online material.}
Besides two-dimensional (2D) cuts and 3D volume rendering,
we follow individual field lines and determine physical properties
along them using the same technique as in \cite{BP11,BP13} and
\cite{CPBC14}.
As an illustration, we show in \Fig{em} the field lines over-plotted with the bright
loop visible in EUV emission at time $t=12.5\,\min$.
Here we display the emission as it would be seen by the Atmospheric
Imaging Assembly \citep{AIA:2012} in its 193\,\AA\ channel showing
basically Fe\,{\sc{xii}} forming at about 1.5\,MK. 
All the field lines are well inside the EUV loop and are therefore
well suited to describing its plasma, magnetic field, and current properties.
The EUV loop is bright only above ${\approx}3$\,Mm because 
of the sensitivity in terms of temperature of the AIA 193\,\AA\ channel.
The currents along these field lines in the loop play a crucial role
for the heating of the loop and its visibility in the EUV \citep[for further
details see][]{CPBC14}.

\section{Results}
\label{S:results}
\subsection{Electric current system in and around the loop}
\label{S:elect.current}
\begin{figure}[t!]
\begin{center}
\includegraphics[width=0.9\columnwidth]{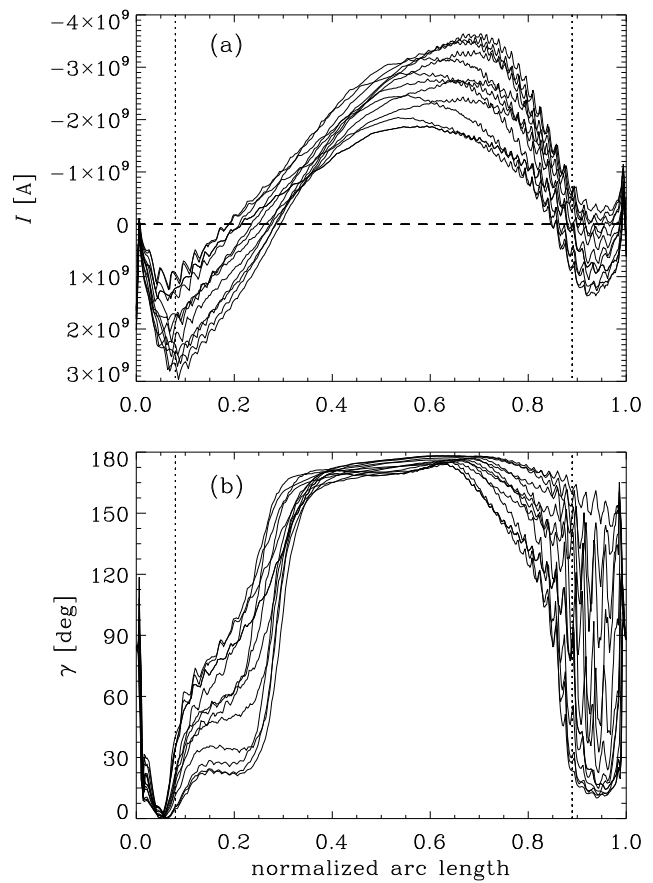}
\end{center}\caption[]{Current properties along a loop in an emerging
  active region of \cite{CPBC14,CPBC15}. The format is the same as for the loop in the stable
  active region in \Fig{sven_totcur}. Panel (a) shows the total
  current $I$ along the loop that now changes sign and panel (b)
  displays the angle $\gamma$ between current and magnetic field. Just
  as in  \Fig{sven_totcur} we show $I$ and $\gamma$ for twelve
  selected field lines in the loop (cf. \Fig{em}) over the normalized
  arc length. Zero arc length is at the left footpoint of the loop
  (see \Fig{feng_vapor}). The two vertical dotted lines indicate the
  coronal region (above 1 MK) and the dashed line indicates zero
  current. We refer to \Sec{S:elect.current}.
}
\label{feng_totcur}
\end{figure}

\begin{figure*}[t!]
\sidecaption
\begin{minipage}[b]{12cm}
\includegraphics[width=12cm]{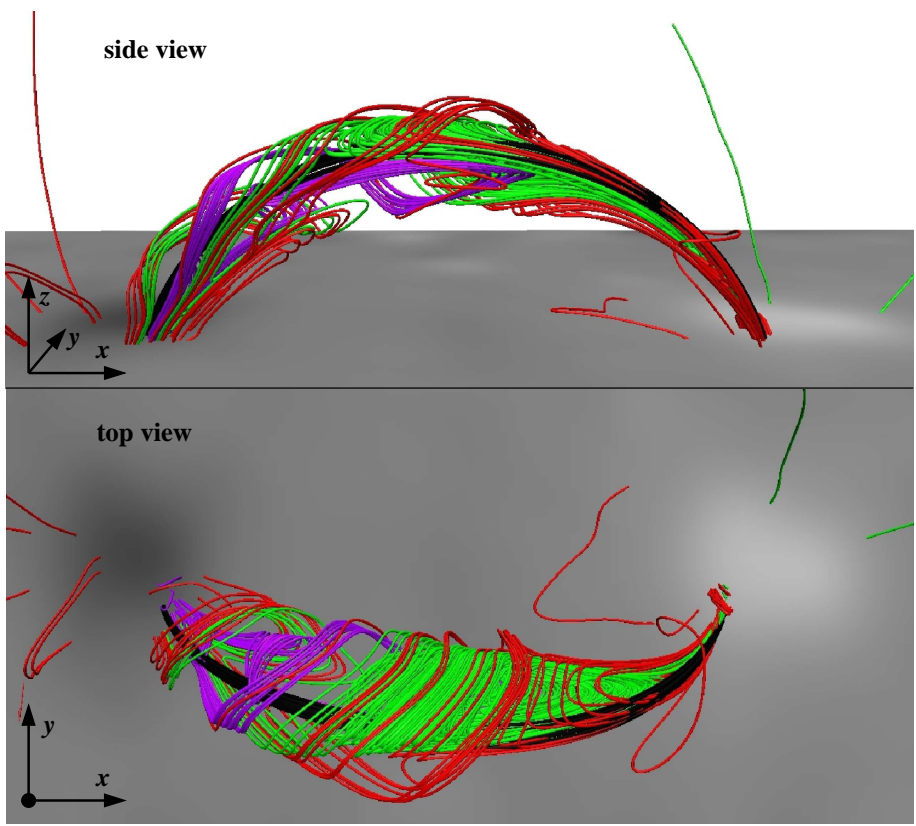}
\end{minipage}
\caption[]{3D rendering of the current and magnetic field lines of the
  loop in the emerging active region of \cite{CPBC14,CPBC15}. The
  two panels show the side view (top panel) and the top view (bottom
  panel) on part of the computational box. The black lines are the field lines within the EUV
  loop (same as in \Fig{em}) and the colored lines show current lines
  (or streamlines of the vector field of the current). The gray scale
  images show the vertical magnetic field at a height of $z=2.9$ Mm at
  the base of the corona where the field already expanded and is
  rather smooth. The three color sets of current lines traced from
  different starting points: Near the positive polarity (red),
  negative polarity  (purple), and from the loop apex (green). We refer to
  \Sec{S:elect.current}. 
}
\label{feng_vapor}
\end{figure*}

We start by analyzing currents along the magnetic field lines in the
loop structure seen in EUV.
For this we define the total current $I$ along the magnetic field $\BB$ as
\begin{equation}
I=\frac{\JJ\cdot\BB}{|\BB|} {\cal A},\quad \text{with}\quad
{\cal A}={|B_z(z=0)|\over |\BB|} \pi r_0^2,
\label{eq:I}
\end{equation}
where $\JJ=\mu_0^{-1}\nab\times\BB$ is the current density and ${\cal
  A}$ is the cross-section of the loop with the constant $r_0{=}0.5$
Mm.
By following an individual field line and normalizing by
the magnetic field this is equivalent to following a flux tube that has
an infinitesimally small diameter at the lower boundary.
Just for convenience we used a value of $2r_0{=}1$\,Mm  at the height
$z{=}0$, that is,\ the bottom boundary in the photosphere.
The loop is then expanding according to flux conservation, that is, the
cross-section ${\cal A}$ increases with decreasing magnetic field
strength $|\BB|$. Basically calculating $I$ through \Eq{eq:I}
corresponds to integrating the current density over the cross-section
of a magnetic flux tube defined by a collection of field lines. The
actual choice of $r_0$ does not matter in our analysis and is only
used for convenience to get sensible units and values for $I$.
A different value of $r_0$ would only result in a different amplitude
of the $I$ but would not give a different dependency along the loop.
 
We now address the question of if and how the total current changes
along the loop in this emerging active region model.  For this we plot
in \Fig{feng_totcur}a the total current $I$ as a function of
normalized arc lengths for twelve different field lines, where the
coronal part is found between the vertical dotted lines. In this
coronal part, the plasma on the respective field line is hotter than
1\,MK. All the field lines show a roughly similar behavior: In about
three quarters of the coronal part the total current $I$ is negative,
whereas in the remaining quarter $ I$ is positive.
This flipping in sign is also consistent with the angle $\gamma$ between
magnetic field and currents, which changes from almost 180$^\circ$ to
30$^\circ$ at the same location (\Fig{feng_totcur}b). At first
glance this sign flip of the currents seems to be counter-intuitive
because it is not consistent with the approximation of a force-free
magnetic field, where the total current has to be  constant and either
parallel or antiparallel throughout the whole loop
(cf. \Fig{sven_totcur}a and \Sec{S:intro}). Thus in this simulation
certainly the magnetic field is not force-free. Of course, we have to
investigate why the magnetic field and current density are parallel on
one side of the loop and antiparallel on the other side.

As a first step, we track the field lines or streamlines of the vector
field of the current density. In the following we  refer to these as
current lines;  they basically track the currents in our system. 
In \Fig{feng_vapor} we show these current lines together with
the magnetic field lines in a 3D volume rendering. 
The magnetic field lines we plot here are confined within the small
cross-section of the actual loop seen in EUV and are the same
field lines as shown in  \Fig{em}. 
These field lines within the EUV loop run more or less parallel without
showing a notable sign of a helical structure. If we were to also plot
field lines further away from the center of the EUV loop, these would
run roughly parallel, too, also not showing clear signs of a helical
structure (we did not show these field lines further out to avoid a
cluttering of the figure). 
The two footpoints of the field lines of the EUV loop are rooted in the
periphery of the magnetic centers, or sunspots, of the active region,
as expected \citep[we refer also to discussion and Fig. 9 in][]{CPBC14}.

In contrast to the magnetic field lines, the overall shape of the
current lines is helical, winding around the magnetic field lines.
To illustrate the overall structure of the current system, we
highlight three types of current lines by color in \Fig{feng_vapor}.
\begin{itemize}
\item
The red current lines are traced starting from the positive magnetic
polarity (right side of \Fig{feng_vapor}) and mostly continuously
connect to the other negative magnetic polarity on the left side.
They illustrate the continuous currents mostly antiparallel to the magnetic
field.
In addition, they show some winding around the magnetic field lines of
the loop.
\item
In contrast, the purple current lines are traced from the negative
magnetic pole on the left side and first closely follow the loop to
the apex. However, following this they change their direction and connect back
to the same negative magnetic polarity on the left side.
\item
The green current lines are traced from the apex of the loop in both
directions towards the surface. These lines are helical and wind
around the magnetic field lines of the loop. Furthermore, there seems
to be more current lines connecting from the apex to the positive
polarity than to the negative one.
\end{itemize}
These cases of current lines illustrate that the current along
one single loop can have both signs, meaning that
currents in the left leg of the loop are pointing
downwards, so parallel to the magnetic field (cf.\ \Fig{feng_totcur}),
and then further up change sign to become antiparallel with the
magnetic field.
This is because currents lines from outside the loop close inside
the loop and thus run in the seemingly wrong direction. Only when
considering the full 3D picture as in \Fig{feng_vapor} can we
understand the seemingly strange change of sign of the current within
the loop as shown in \Fig{feng_totcur}a.
In their model, \cite{AH14}, investigate the injection of a
magnetic flux sheet into a convection-driven model of the solar corona.
They also find a complex current structure
around a smooth magnetic field and in their model the currents are
able to trigger several small flares.

Furthermore, the complex helical structure of the current lines
in our work suggest that the force-freeness of the magnetic field is only
partially fulfilled.
This is true not only in the surroundings of the loop as illustrated
by the current lines in \Fig{feng_vapor}, but also within the bright
EUV loop, where in particular in the left quarter of the coronal part
(from 0.1 to 0.3 normalized arc length) the angle between magnetic
field and current is of the order of 30$^\circ$, so certainly not
(anti-)parallel.

\subsection{Driving at the base of the corona}
\label{S:driv.cor}

\begin{figure}[t!]
\begin{center}
\includegraphics[width=\columnwidth]{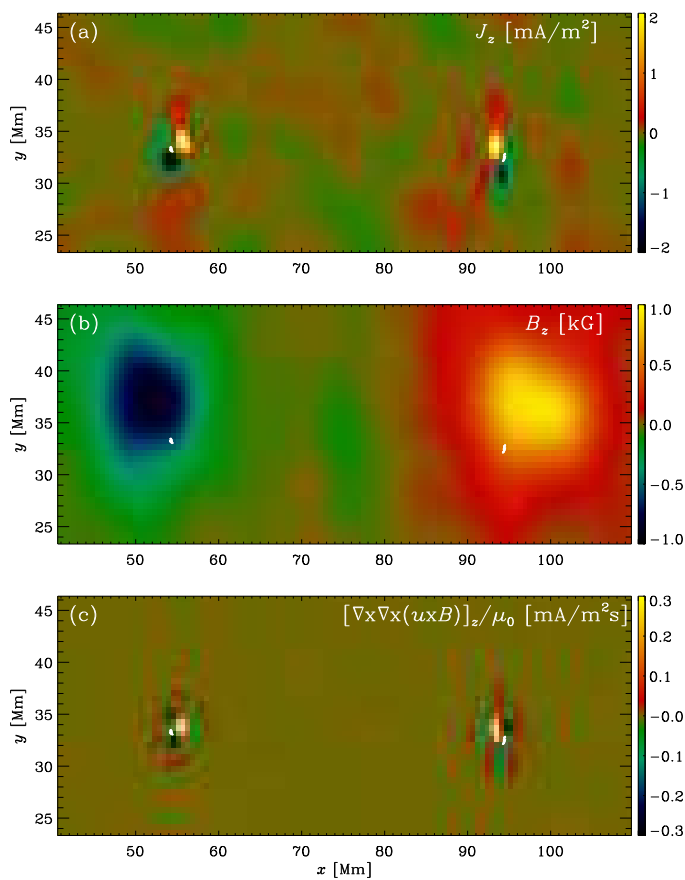}
\end{center}\caption[]{Conditions at the base of the corona of \cite{CPBC14,CPBC15}.
Vertical current density $J_z$ (a), vertical magnetic field $B_z$
(b) and the vertical component of the double curl of the electromotive force
$[\nab\times\nab\times(\uu\times\BB)]_z$ (c)
in a horizontal plane at a height of $z=2.9\ $Mm placed around the
field lines (white dots); we refer to
\Sec{S:driv.cor}.
}
\label{feng_jz}
\end{figure}

What drives the currents within the loops system to change sign? To answer
this we investigate the currents and the driving forces at the base of
the corona.  The flip of sign of the currents with respect to the
magnetic field (as seen in \Fig{feng_totcur}a) implies that on both
sides the currents are downward directed.
On the left footpoint the magnetic field is negative (see
\Fig{feng_jz}b), thus downward directed. There the angle $\gamma$
between magnetic field $\BB$ and current is small (\Fig{em}), so the
currents are parallel to $\BB$ and thus downwards. At the right
footpoint the magnetic field is positive (upwards) but the angle
$\gamma$ is about 180$^\circ$, so the current is antiparallel to $\BB$
and thus also downwards directed.
This is also the case at the base of the corona. To illustrate this
we show in  \Fig{feng_jz}a the vertical current at a constant height of 2.9\,Mm. 
At the location where the field lines penetrate this  height
layer, the vertical currents at both loop footpoints are
negative. That the downward directed currents at both footpoints are
present at the base of the corona suggests that the direction of the
currents in the loop is set by the dynamics at the bottom of the
loop.
An interesting feature is the bipolar curl structure surrounding the
loop; in particular, it is significantly stronger than the overall
current at this height (see \Fig{feng_jz}a).
Directly in the neighborhood of the loop footpoint, the currents are
negative, whereas towards the center of the plane (away from the
sunspot), the currents are positive.
Because of the symmetry of the simulation, the bipolar vertical
current structure near the two sunspots is symmetric (see
\Fig{feng_jz}a-b).

The currents at the base of the corona are generated by the velocity
acting on the magnetic field.
This can be seen by taking the curl of the induction equation,
which describes the evolution of the current density.
On the right-hand side (rhs), the dominant term is the double curl of the electromotive
force $\nab\times\nab\times(\uu\times\BB)$, whose vertical component
is plotted in \Fig{feng_jz}c.
It shows clearly that the sign of current is generated by velocity
field, which is driven by the convective motions in the
photosphere that move and shear the magnetic field.
However, the values indicate a typical timescale of just 10 seconds to
produce the amount of current density seen in \Fig{feng_jz}a.

Because of the small-scale flows in the photosphere and the resulting
fluctuation of the (vertical) currents, it is difficult to pin-point
the exact cause of the bipolar vertical current structure at the base
of the corona (at a height of 2.9 Mm as visualized in
\Fig{feng_jz}a).
However, we see upflows into the loop and a coalescent flow driving
the loop footpoints into the sunspots \citep[as already pointed out
by][]{CPBC15}.
In accordance with the above discussion of the double-curled
electromotive force (\Fig{feng_jz}c), this flow structure is a good
candidate for understanding the formation of the bipolar current structure
at the loop footpoints.

In addition, we attempt to describe the current in the uncurled
induction equation, where they are related to difference of (1) the electromotive
force $\uu\times\BB$ and (2) the temporal evolution of the
magnetic field or vector potential, respectively.
The second is governed by the magnetic field evolution enforced at the photospheric level,
that is,\ the magnetic field in the form of flux tubes is pushed through the lower boundary.
A detailed analysis reveals that from the photosphere to the base of
the corona these two contributions, that is, the electromotive force
and the time evolution, are similar in
strength and structure resulting in a small difference, which is\
the curl of the current.
A clear correlation between $\uu\times\BB$ and the vertical
component of the current is not present, however. 
This is not
surprising because the currents are the (small) difference of two
larger quantities. The typical structure of the currents at the base
of each loop footpoint seen at  $z=2.9\Mm$ in \Fig{feng_jz}a is
similar for heights near the coronal base.
However, closer to the photosphere, the structure of the vertical
currents becomes more complex, which is because of the small
convection patterns. This renders the analysis of the ultimate origin
of the sign of vertical current using the curled and uncurled
induction equation in the photosphere an ill-posed problem.

\begin{figure*}[t!]
\sidecaption
\begin{minipage}[b]{12cm}
\includegraphics[width=12cm]{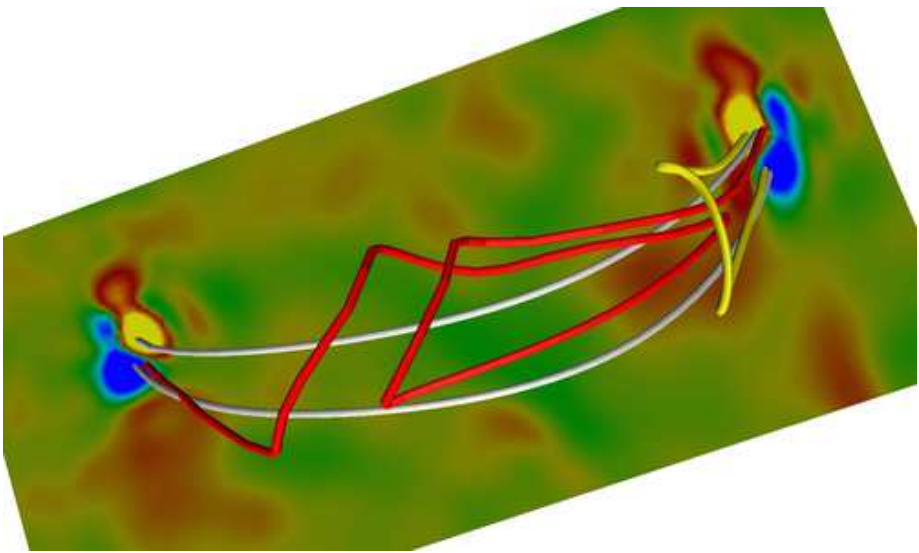}
\end{minipage}
\caption[]{3D rendering of the current and magnetic field lines around the
  loop in a top view from above the active region of \cite{CPBC14,CPBC15}. The colored
  plane at the bottom shows the vertical current $J_z$ at $z=2.9$ Mm at
  the base of the corona; the color-coding and scale are similar to those in
  \Fig{feng_jz}a for the same quantity at the same height. The gray-colored lines display the magnetic field lines connecting the
  positive and negative vertical current concentrations with each
  other. The red line shows a current line (or a
  streamline of the vector field of the current) connecting the
  positive vertical current concentration on the right-hand side to
  the negative concentration on the left-hand side. The yellow line also
  shows a current line, but connecting the positive and negative
  current concentrations on the right-hand side. 
We refer to \Sec{S:driv.cor}.
}
\label{jz_vapor}
\end{figure*}

The currents in the corona, and in the loop in particular,
simply adjust to the current structure at the base of the corona.
In particular, the bipolar current structure in \Fig{feng_jz}a plays an
important role in forming the current structure around and in the
loop.
To illustrate this fact, we show in \Fig{jz_vapor} only a few selected magnetic
field lines (gray) and current lines (red, yellow) connecting the bipolar vertical current
concentration at the coronal base.
One magnetic field line connects the positive (vertical) current on
the left side to the positive current on the right side. The other
fieldline connects the two patches of negative currents. So, at each of
the two footpoints of each fieldline, the vertical current points in
the same vertical direction (both times in or out of the loop). This
is not a feature of these two selected fieldlines, but a general
feature and reflects the discussion of the currents along selected
fieldlines in \Sec{S:elect.current}.

The current lines behave differently.
One current line (red) starts from the positive concentration on the
right-hand-side, close to one of the field lines and ends at the
negative concentration of the left-hand-side close to other magnetic field lines. 
On its way from one side to the other it leaves the magnetic field
line that it started with, winds around the magnetic field (indicating a
circular component of the current) and ends up with the another magnetic
field line.
The other current line (yellow) starts on the right side not too far
from the other current line at the positive concentration, but
connects to the negative current concentration on the same right side
close to the other field line.
This clearly shows that the currents along magnetic field lines change
their sign, because the current lines have to connect the two
different current polarities and therefore cannot be aligned with the
magnetic field.
Furthermore, the bipolar current structure is crucial for the current
field line to wind around the loop or even turn back to the same side
of the loop.
However, in \Sec{S:flows.mag} we discuss that the plasma flows within
the loop also have an influence on the helical loop structure.

\subsection{Induced magnetic field and the coronal currents}
\label{S:consequences}

\begin{figure}[t!]
\begin{center}
\includegraphics[width=\columnwidth]{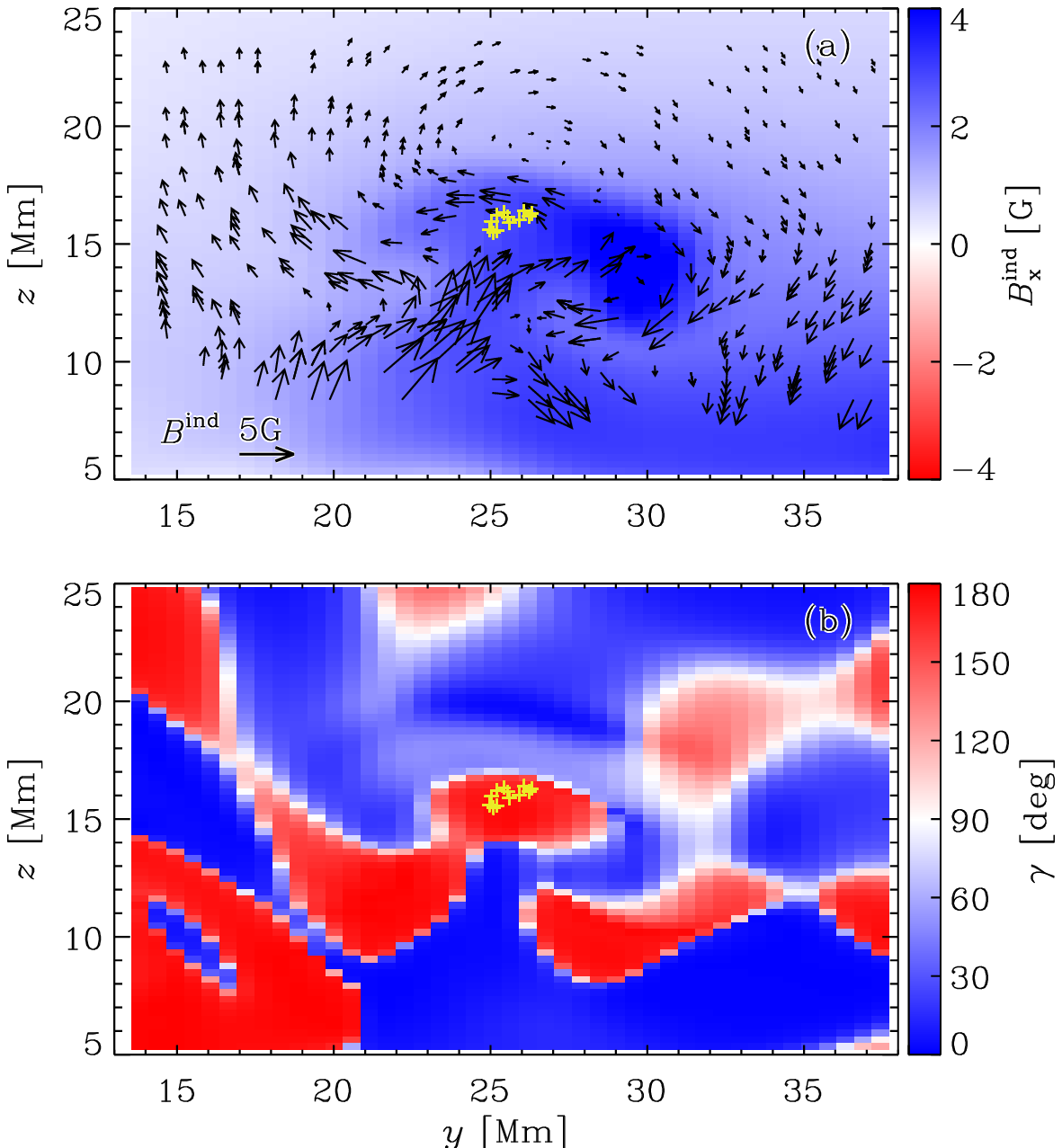}
\end{center}\caption[]{Magnetic field strength and inclination at the
  apex of the loop at $x=75\,$Mm of \cite{CPBC14,CPBC15}.
(a) Induced magnetic field: color-coded $\Bind_x$ together with
magnetic field vectors $\Bind$ in $yz$ plane.
We note that the positive $x$ direction is pointing out of the plane of view.
(b) Angle $\gamma$ between current density $\JJ$ and the magnetic
field $\BB$, see \Eq{eq:gam}.
The locations, where the magnetic field lines \Fig{feng_totcur} breach
through the plane are marked with yellow crosses. We refer to \Sec{S:consequences}.
}
\label{bfield}
\end{figure}

The complex helical current system around the bright loop structure is
a consequence of the magnetic field structure and the forces associated with it.
The large-scale current system in the corona associated with the loop
turns out to be strongly influenced by induced coronal magnetic field
that locally reduces the background magnetic field of the active
region.

As a first step, we investigate how the magnetic field deviates
from a current-free magnetic field, that is, a potential field.
For this purpose, we extrapolate a potential field using the
vertical magnetic field at the bottom boundary ($z{=}0$).
We identify the difference of the potential and the actual magnetic
field as the induced magnetic field $\BBind$. This is the component of
the magnetic field associated with the currents. To illustrate the
induced field we show a vertical cut through the mid-plane between the
two loop footpoints ($y$-$z$ plane at $x{=}75\,$Mm) in \Fig{bfield}a.
Near the loop apex (at $z{\approx}15$\,Mm) $\Bind_x$ has up to 4\,G in
the positive $x$ direction.
Because the overall magnetic field $\BB$ points in the negative $x$
direction at the apex with a strength of around 100 G, the induced
field $\BBind$ reduces the background magnetic field by around 5$\%$.
The induced magnetic field in \Fig{bfield}a has a shape similar to a
tilted mushroom \citep[also visible in Fig. 8 of][]{CPBC14}.
The induced field vectors in the vertical cut form a left-handed helical field
above and beside the loop, where the field strength is smaller than in
the $x$ direction (\Fig{bfield}a).
Below the loop, there exists a strong vertically upward-pointing induced magnetic
field with a comparable strength to $\Bind_x$.
In essence, while the overall magnetic field is running more or less
straight from one footpoint to the other, this  shows that the induced
field has a clear helical structure. Nevertheless, because the induced field
has a strength of only 5\% of the overall background field, the
helicity is not visible when simply plotting magnetic field lines.

\begin{figure}[t!]
\begin{center}
\includegraphics[width=\columnwidth]{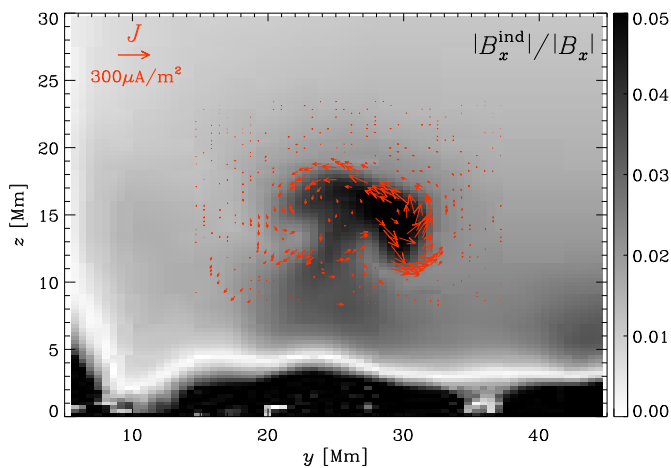}
\end{center}\caption[]{Relative induced magnetic field and current
  density in the $y$-$z$ plane of the loop apex at $x=75\,$Mm of \cite{CPBC14,CPBC15}.
The ratio of the modulus of the induced magnetic field in the $x$
direction $|\Bind_x|$  and the modulus of the total magnetic field $|B_x| $ is color-coded.
The red arrows indicate vectors of the current density $\JJ$ in the $y$-$z$
plane. We refer to \Sec{S:consequences}.
}
\label{bind}
\end{figure}

To check that the induced magnetic field is consistent with the currents,
we plot these together in \Fig{bind}. Here we see the mushroom-type enhancement
of the induced field that is restricted to the vicinity of the loop. While
the actual EUV loop visible in coronal emission has a diameter of some  2\,Mm
\citep{CPBC14}, here the region of the induced magnetic field covers a region
(in a vertical cut near the apex) of almost $10{\times}5$\,Mm$^2$. While
in this region the induced field is of the order of 5\% of the background
field, it is practically zero outside in the corona (of course,  in the photo-
and chromosphere where the field is far from being force-free the induced
magnetic field is much larger). The currents in the vertical cut in \Fig{bind}
show a clear counter-clockwise rotation, which according to the right-hand-rule
is directly related to induced magnetic field in the positive $x$
direction covering an area much larger than the EUV loop.

To relate the induced magnetic field to the current system around the
loop, we define the angle $\gamma$ between current and magnetic field
so that
\begin{equation}
\cos{\gamma}=\frac{\JJ\cdot\BB}{|\JJ|\,|\BB|}~.
\label{eq:gam}
\end{equation}
This allows us to quantify the discussion of the helical currents
around the loop as visualized in \Fig{feng_vapor}. To this end we plot
the angle $\gamma$ in \Fig{bfield}b in the same vertical mid-plane between
the footpoints as in \Fig{bfield}a. 
In accordance with  \Fig{feng_totcur}  we see that at
the position of the loop (indicated by the yellow crosses), the
currents are antiparallel to the field
($\gamma{\approx}180^\circ$). In contrast,  away from the center of
the loop the magnetic field deviates significantly from a force-free
state as emphasized by values of $\gamma$  close to 90$^\circ$ with
currents being almost perpendicular to the magnetic field, in
particular in a large patch directly above the loop (see
\Fig{bfield}b). 
This consideration clarifies that the strongest deviation from a
force-free field is found  outside the loop. One might be tempted to
conclude that the field is force-free inside the loop, where the EUV
emission is strong. However, this would not be correct. While there
the currents are indeed parallel or antiparallel to the field (consistent
with force-free) the currents can switch sign in response to the
closing of the current system surrounding the loop
(cf.\,\Sec{S:elect.current}), which cannot be captured by a force-free
description of the magnetic field.

We summarize that in a region significantly larger than the coronal loop
seen in EUV we find a significant disturbance of the magnetic field
directly related to the current systems.
Inside and outside the EUV loop the assumption of a force-free field
breaks down.

\begin{figure}[t!]
\begin{center}
\includegraphics[width=0.98\columnwidth]{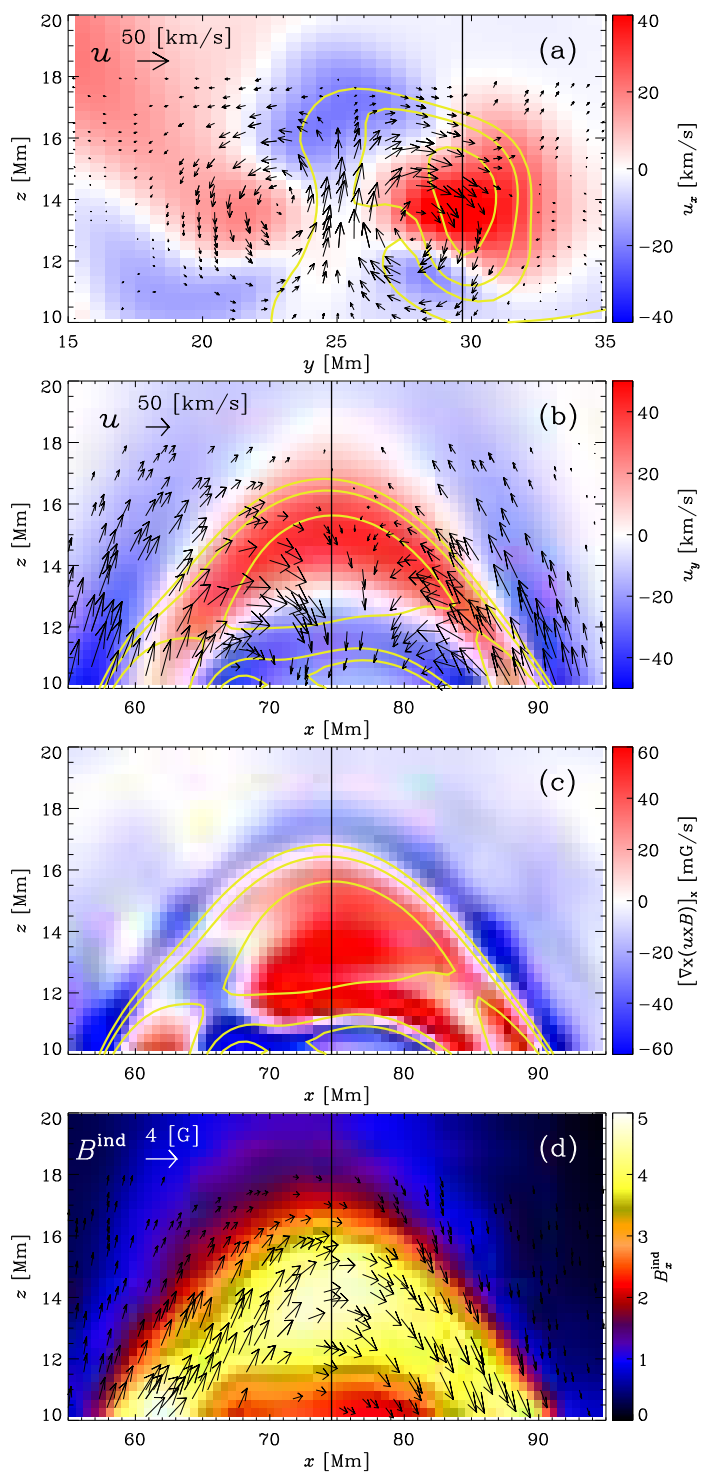}
\end{center}\caption[]{Flows, curl of electromotive force, and induced magnetic
  field in the loop structure of \cite{CPBC14,CPBC15}. The vertical cuts are perpendicular to
  the loop in the $y$-$z$ plane  at $x\approx$75\,Mm (a) and along the
  loop in the $x$-$z$ plane  at $y\approx$30\,Mm (b--d) as indicated
  by the vertical lines in the corresponding cuts. Panels (a,b)
  show the velocity $\uu$ in, the arrows indicating the components
  within the plane, and
  the color of the component out of the plane. Panel (c) shows  curl of the electromotive force in the $x$ direction
  $[\nab\times(\uu\times\BB)]_x$ color-coded in the $x$-$z$ plane. Panel
  (d) displays the induced magnetic field $\BB^{\rm ind}$, color-coded for the $x$
  component and the arrows  showing the vectors in the $x$-$z$ plane.
The yellow contours in panels (a) to (c) indicate the levels of
$2.5,3,4$ G for $B_x^{\rm ind}$.
We refer to \Sec{S:flows.mag}.
}
\vspace{-0.5cm}
\label{uxbb}
\end{figure}

\begin{figure}[t!]
\begin{center}
\includegraphics[width=\columnwidth]{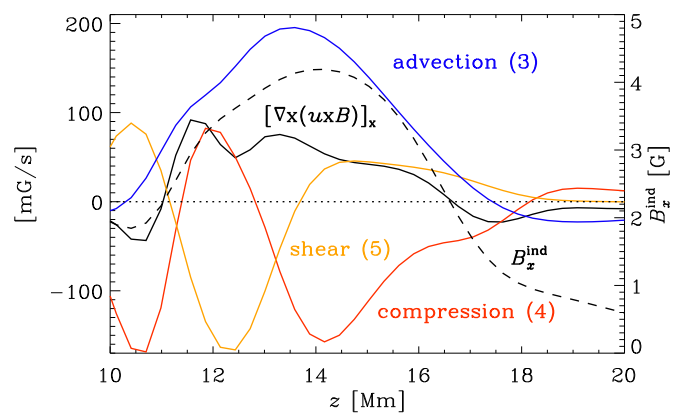}
\end{center}\caption[]{$x$ component of the curl of electromotive force
  $[\nab\times(\uu\times\BB)]_x$ (black
  solid line) together with its main contributions, advection
  (blue), compression (red) and shear (orange) plotted over height $z$
  in the location of high induced magnetic field ($x=75\,$Mm, $y=30\,$Mm, also
  indicated in \Fig{uxbb}a and c as vertical lines); we refer also to \Eqss{eq_uxb1}{eq_uxb4} for the
  contributions. Additionally we over-plot the same
  component of induced magnetic field $B_x^{\rm ind}$ (dashed
  black). The horizontal black dotted line represents the
  zero value. Data from of \cite{CPBC14,CPBC15}. We refer to \Sec{S:flows.mag}.
}
\label{uxbc}
\end{figure}

\subsection{Plasma flows and induced magnetic field}
\label{S:flows.mag}

To understand the origin of the induced magnetic field, we investigate the
plasma motions in the loop structure. 
These motions can generate a (small) change in the magnetic field
through the electromotive force.
The plasma flows from the bottom of the corona into the loop are shown in \Fig{uxbb}a and b, where we plot the
plasma velocities in the $y$-$z$ and $x$-$z$ planes, that is, in vertical planes across and along the loop.
The flows are mostly upwards in both legs of the loop (\Fig{uxbb}b).
The cut perpendicular to the loop in the $y$-$z$ plane reveals that
the plasma moves upwards in the middle of the loop and disperses in
the top of the loop in horizontal directions (\Fig{uxbb}a).
Transported away from the loop the plasma falls down again generating
a small vortex-like flow structure in the $y$-$z$ plane together with a
velocity in the $x$ direction. 
In one of these vortex-like structures the $x$-component of the induced magnetic field has a
maximum, indicated with the yellow contours in \Fig{uxbb}a.
A vertical cut through this location at 29.5\,Mm (\Fig{uxbb}b) shows the upflows along
the legs of the loop, the horizontal flow at top of the loop, and the
downflows below the apex.
The upward motion along the legs is consistent with the analysis by \cite{CPBC14}, where the flow is found to be driven by the Joule heating
at the base of the corona.

The presence of the flow vortex and the induced magnetic field at the
same location (\Fig{uxbb}a) indicates a connection between these two.
To check this, we computed the curl of the electromotive force in the $x$
direction because this is the main contribution for changing $B_x$ in time.
We find positive values for the electromotive force of some 20\,mG/s to
60\,mG/s (in the positive $x$ direction) at similar locations where
also the large values of induced magnetic field are found
(cf. \Fig{uxbb}c).
In particular, this is also the case at the same height as the loop
and slightly below. Analyzing the same simulation, \cite{CPBC14} found
that the flows into the loop last some 100\,s (their
Fig.\,3d). Consequently, together with the electromotive force, over
this timescale we can expect an induced magnetic field of some 2\,G
to 6\,G. This is just the induced magnetic field we find
(cf.\Fig{uxbb}d), that also points in the positive $x$ direction. This
induced magnetic field then slightly suppresses the background magnetic
field, which is pointing in the negative $x$ direction near the loop
top. So from this we can conclude that the flow in the loop induces a
magnetic field that is directed opposite to the background field and
thus reduces the field strength in the loop.

The induced magnetic field also has components in the other directions (\Fig{uxbb}d),
but we focus here on the $x$ component as it is tightly related to the helical
currents around the loop; we refer to \Sec{S:consequences}.
However, the other components of the induced magnetic field can also
be linked in a similar way to their corresponding components of the
curl of electromotive force.

As a next step we want to investigate which contribution of the flow acting on the magnetic field is dominant.
The $x$ component of curl of the electromotive force can be
divided into four terms;
\begin{eqnarray}
&&[\nab\times(\uu\times\BB)]_x = \nonumber\\
\label{eq_uxb1}
 &&-u_y\partial B_x/\partial y-u_z\partial B_x/\partial z\quad \quad \text{advection}\\
\label{eq_uxb2}
 &&-B_x\partial u_y/\partial y-B_x\partial u_z/\partial z\quad \quad \text{compression}\\
\label{eq_uxb3}
 &&+B_y\partial u_x/\partial y+B_z\partial u_x/\partial z\quad \quad \text{shear}\\
\label{eq_uxb4}
 &&+u_x\partial B_y/\partial y+u_x\partial B_z/\partial z\quad \quad
 \text{solenoidality of } \BB.
\label{eq_uxb5}
\end{eqnarray}
We note that the terms $u_x\partial B_x/\partial x$ and
$B_x\partial u_x/\partial x$ are often added for completeness, but
because their contributions cancel out, we do not show them here.
Furthermore, we prefer this kind of separation because it makes it
easier to disentangle the different contributions.
The last contribution turns out be small and is neglected in
the following discussion.
In \Fig{uxbc}, we plot the first three contributions over height at a
location, where $B^{\rm ind}_x$ is large in the middle of the loop
($x=75\,$Mm, $y=30\,$Mm, also indicated in \Fig{uxbb}a and c as
  black vertical lines).
Again, a positive electromotive force coincides with large values of
$B_x^{\rm ind}$.
This is true between $11$ and $17\,$ Mm in height. Below and above
$[\nab\times(\uu\times\BB)]_x$ is negative and $B^{\rm ind}_x$ small.
The positive values of the curl of electromotive force can be mostly associated
with the advection (\ref{eq_uxb1}) of magnetic field along the plasma.
Between $z=14$ and $17$ the contribution of the shear (\ref{eq_uxb3}) and between
$z=11$ and $13$ the contribution of the compression (\ref{eq_uxb2}) is also positive,
but the advection (\ref{eq_uxb1}) always dominates at these heights.
A more detailed analysis reveals that compression becomes dominating in
the legs.
It is clear that considering all contributions is important to obtaining a full and accurate picture of magnetic field evolution.
Pure advection would result in an overly large change in the magnetic
field, therefore only fully compressible 3D MHD simulations are able
to capture these plasma forces
acting on the magnetic field and leading to the observed helical
current structure.

\begin{figure}[t!]
\begin{center}
\includegraphics[width=\columnwidth]{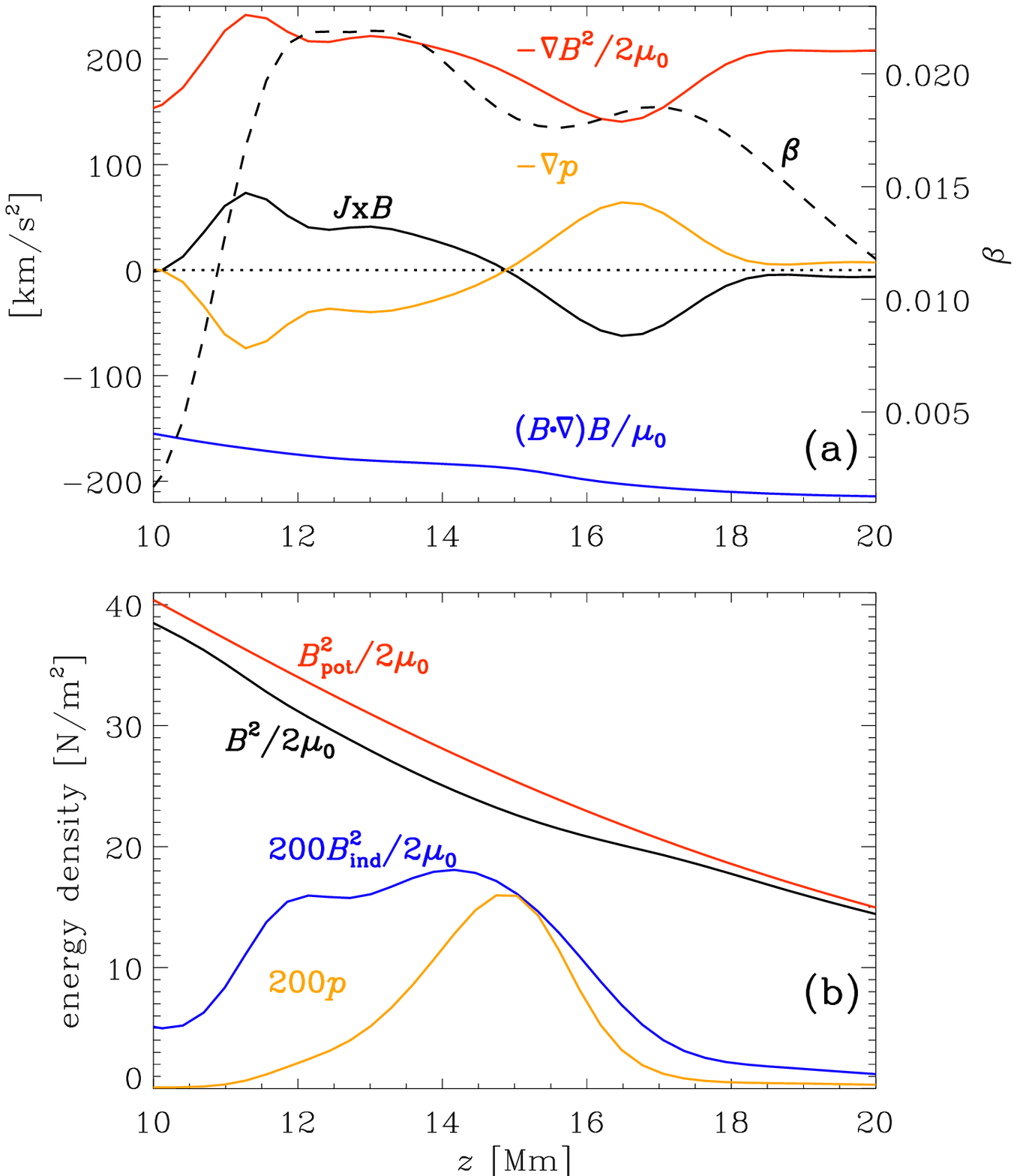}
\end{center}\caption[]{Vertical force balance and energy density in
  the vicinity of the loop of \cite{CPBC14,CPBC15}. 
(a) Vertical force balance in terms of acceleration with
  contribution from the pressure gradient $\nab p$ (orange solid line)
  and the Lorentz force $\JJ\times\BB$ (black) divided in the magnetic
  pressure gradient $\nab B^2/2\mu_0$ (red) and the magnetic tension
  force $(\BB\cdot\nab)\BB/\mu_0$ (blue) plotted over height in the vicinity
  of the loop ($x=75\,$Mm, $y=30\,$Mm).
We also show the plasma $\beta$ for the same region (black dashed)
($y$-axis on right-hand side).
(b) Energy densities of the total magnetic field $B^2/2\mu_0$ (black line),
potential magnetic field $B_{\rm pot}^2/2\mu_0$ and the induced magnetic
field $B_{\rm ind}^2/2\mu_0$ shown together with the thermal pressure
$p$ (orange).
The horizontal black dotted line in the upper plot represents the zero
value. We refer to \Sec{S:flows.mag}.
}
\label{forc}
\end{figure}

The prominent flows and the induced magnetic field have an
additional effect on the plasma and the force balance of the system.
As shown in Fig. 8c of \cite{CPBC14}, the loop shows a concentration of
density in the top-right part of the mushroom-type structure seen in
the cross-sectional cut of the loop.
This coincides well with the location of the induced magnetic field
shown here in Figs. \ref{bfield}a and \ref{uxbb}a.
This local density enhancement can be explained by the upflows that
transport  material from the base of the corona into the
loop structure.
The density  remains high there, because the magnetic field can keep it
at the same position.
The enhanced density leads to an enhanced pressure; the temperature is
more uniformly distributed; we refer to Fig 8a of \cite{CPBC14}.
In \Fig{forc}, we plot the force balance and the distribution of
pressure and magnetic energy over height at the same location as in
\Fig{uxbc}.
The vertical force due to the pressure is exactly balanced by the
magnetic pressure component of the Lorentz force (\Fig{forc}a).
The induced magnetic field lowers the background magnetic field strength and therefore
the magnetic pressure in this region.
This allows the generation of a local enhancement of the gas pressure.
As plotted in \Fig{forc}b, the gas pressure is enhanced in the same
region, where the induced magnetic field is present.
Even though the plasma $\beta$ is very small ($\beta\approx
0.018$)\footnote{We note that the difference between the value of
  $\beta$ and the actual ratio of $p$ and $B^2/2\mu_0$ is due to the
  applied factor reducing the Lorentz-force, in the cases when the
  Alvf\'en speed becomes too large; we refer to Section 2 of \cite{CP15}.}
in this region, the gradient of pressure balances the Lorentz force.
This seems to be counter-intuitive.
However, the parts of the magnetic field and the pressure
that do not change in space do not contribute to the force balance, but they define the value of
$\beta$.
Therefore, the value of plasma $\beta$ in this region is misleading
when investigating the force balance that is governed by the
gradients. 

\subsection{Squashing of the loop}
\label{S:squashing}

The plasma flows into and through the loop structure have  consequences for the
evolution of the magnetic field.
\cite{CPBC14} reported a squashing of the loop in the later evolution.
They suggested, that the interaction of the magnetic field of
the loop with overlying coronal magnetic field is responsible for
compression and squashing of the loop.
A detailed analysis reveals that this conclusion might not be correct.
The flows shown in \Fig{uxbb}a and b act also on the magnetic field
directly above the loop.
As shown in the \Fig{uxbb}c, the resulting change in magnetic field
through the electromotive force is negative above the loop structure
and positive inside the loop.
Even though this plot is a cut through the right part of the
mushroom-type structure (cf. \Fig{bind}), the values are similar to those in the middle part of the loop.
Because the background magnetic field points in the negative $x$ direction,
outside (above) the loop the flow results in an amplification, inside the loop to a suppression of the magnetic field. 
The amplitudes of the resulting change in the magnetic field are
around $50\,$mG/s.

This change of the magnetic field induced by the flow in the loop
might well account for the squashing of the magnetic structure
reported by \citet{CPBC14}. They found that the squashing of the loop
takes place over the course of about 500\, to 1000\,s. During this
time the electromotive force (induced by the flow) of about 50 mG/s
will change the magnetic field by some 25\,G to 50\,G.
This is a significant fraction of the background magnetic field
strength, which is about 100\,G near the loop apex.
Above the loop the background magnetic field gets significantly enhanced, within the loop reduced. 
Essentially this looks like a pile-up of magnetic field above the
expanding loop, or like a squashing of the loop.

We did not find any evidence that the overlying magnetic field plays
an important role in the squashing process.
Its contribution would show up in the diffusion term of the induction
equation, but this term is one order of magnitude smaller
than the flow acting on the magnetic field. 
Therefore we conclude that the expanding loop gets squashed through a
change of magnetic field induced by the plasma flow through the loop.

\section{Conclusions}\label{S:conclusions}

In this work we analyzed the magnetic field and current structure of a
bright coronal loop of the three-dimensional MHD simulation of
\cite{CPBC14}.
We found that while the magnetic field lines follow the loop structure
as a concentrated flux bundle, the current stream lines form a complex
structure.
In particular, in the coronal part of the loop the current density undergoes a change
of direction along the loop.
This behavior is related to the footpoint motions in the
photosphere, where a bipolar vertical current structure is generated.
Because the magnetic field connects smoothly the current concentrations with the
same vertical direction, the currents cannot be aligned with the
magnetic field.
Furthermore, the bipolar current structures play an important role in
forming a complex current system because the current stream lines have to
wind around the loop to connect to the bipolar current concentration on
the other side of the loop.

Most of the coronal part of the simulation is close to a current-free
magnetic field configuration.
We find that the currents are directly related to a reduction of  the
background magnetic field by about 4$\%$, forming an isolated non force-free structure.
Because this structure has the same location as the hot bright coronal
loop, a non-force-free state seems to be a necessity for heating the
loop to coronal temperatures.

The local reduction of the background magnetic field can be explained
by plasma flows into and through the loop acting on the magnetic
field. In agreement with \cite{CPBC14}, we found upward flows along
the legs. The upward directed expansion of the loop creates vertical
vortices next to the loop. This flow structure can be directly linked
to the reduction of the magnetic field due to the curl of the
electromotive force. The main contribution is the advection of field,
but also compression and shear play a role.

The plasma flows acting on the magnetic field reduce the magnetic
pressure. This allows for the stable concentration of density inside
the loop, because the pressure gradient is balanced by the Lorentz force.
This enhancement of density is crucial for the emission from the loop seen in extreme UV wavelengths.
Additionally, we can relate the flow structure to the magnetic field
evolution and explain the  squashing of the loop at later times.

This work shows that a realistic modeling of a coronal loop
structure, which produces realistic emissivity in coronal lines,
needs to resolve the three dimensional structure of complex magnetic
field and current systems.
Even in a low-$\beta$ environment, the plasma properties, that is, flow,
density, and pressure, will significantly influence the magnetic field structure and evolution.
This is in particular true if the coronal loop is formed above an emerging active region.
Therefore, one dimensional loop models \citep[e.g.,][]{Klim06},
force-free extrapolations \citep{Wi08} and magneto-frictional models
\citep[e.g.,][]{CD12} are most likely not able to reproduce the complex
helical current structure and the related UV emission as found in our
models. 

The model analyzed here, of course, is not a perfect
representation of an emerging active region on the Sun.
The coronal model discussed here builds on a model for the emergence of magnetic flux from the upper
convection zone to the surface \citep{RC14}.
On one hand, the flux emergence in this model, as well as in the related model of
\cite{Cheung_etal10}  is significantly faster than in the real Sun
\citep{BSBCGLR16}.
Therefore, the changes in our modeled coronal magnetic field might be
faster than on the Sun.
On the other hand, the  flux tube emerging in our model is not twisted
\citep[in contrast to the model of ][]{Cheung_etal10}. A flux tube
with a significant twist breaking through the surface  would lead to a significant
amount of magnetic helicity being injected into the corona and thus
could lead to  highly twisted magnetic structures in the corona. In
that sense the considerations presented here represent lower limits
for the amount of complexity of the current structures in and around
coronal loops.

Further studies will involve the investigation of key parameters responsible
for complex current systems.
One possibility could be indeed the magnetic helicity, which is known to play
an important role in the formation of unstable loop structures
\citep[e.g.,][]{TK05,PVDD15} and is connected to the underlying dynamo
\citep[e.g.,][]{WB10,WBM11,WKMB12}.
Another future work path would be to perform realistic coronal simulations on top
of self-consistently formed bipolar flux concentration, either in convection
\citep{SN12,KBKKR16} or forced turbulence \citep{WLBKR13,WLBKR16}.

\begin{acknowledgements}
The authors thank the anonymous referee and Matthias Rempel for useful
comments and suggestions.
The simulations have been carried out on supercomputers at
Gesellschaft f\"ur wissenschaftliche Datenverarbeitung mbH
G\"ottingen (GWDG), and on the supercomputers at Max Planck Computing
and Data Facility (MPCDF) in Garching and SuperMUC. We thank the
Partnership for Advanced Computing in Europe (PRACE) for awarding us the access to SuperMUC
based in Germany at the Leibniz Supercomputing Center (LRZ).
J.W. acknowledge funding by the Max-Planck/Princeton Center for
Plasma Physics and the funding from the People Programme (Marie Curie
Actions) of the European Union's Seventh Framework Programme
(FP7/2007-2013) under REA grant agreement No.\ 623609.
This work was partially supported by the International Max-Planck
Research School (IMPRS) on Physical Processes in the Solar System and
Beyond (F.C.).
The National Center for Atmospheric Research is sponsored by the
National Science Foundation.
The visualizations of \Figs{feng_vapor}{jz_vapor} were done using {\sc
  VAPOR}\footnote{{\tt www.vapor.ucar.edu}}.
\end{acknowledgements}

\bibliographystyle{aa}
\bibliography{paper}

\begin{thebibliography}{38}
\expandafter\ifx\csname natexlab\endcsname\relax\def\natexlab#1{#1}\fi

\bibitem[{{Archontis} \& {Hansteen}(2014)}]{AH14}
{Archontis}, V. \& {Hansteen}, V. 2014, \apjl, 788, L2

\bibitem[{Bingert(2009)}]{Bingert2009}
Bingert, S. 2009, PhD thesis, Albert-Ludwig-Universit{\"a}t Freiburg

\bibitem[{{Bingert} \& {Peter}(2011)}]{BP11}
{Bingert}, S. \& {Peter}, H. 2011, \aap, 530, A112

\bibitem[{{Bingert} \& {Peter}(2013)}]{BP13}
{Bingert}, S. \& {Peter}, H. 2013, \aap, 550, A30

\bibitem[{Birch {et~al.}(2016)Birch, Schunker, Braun, Cameron, Gizon,
  L{\"o}ptien, \& Rempel}]{BSBCGLR16}
Birch, A.~C., Schunker, H., Braun, D.~C., {et~al.} 2016, Science Advances, 2

\bibitem[{{Boerner} {et~al.}(2012){Boerner}, {Edwards}, {Lemen}, {Rausch},
  {Schrijver}, {Shine}, {Shing}, {Stern}, {Tarbell}, {Title}, {Wolfson},
  {Soufli}, {Spiller}, {Gullikson}, {McKenzie}, {Windt}, {Golub}, {Podgorski},
  {Testa}, \& {Weber}}]{AIA:2012}
{Boerner}, P., {Edwards}, C., {Lemen}, J., {et~al.} 2012, \solphys, 275, 41

\bibitem[{{Bourdin} {et~al.}(2013){Bourdin}, {Bingert}, \& {Peter}}]{BBP13}
{Bourdin}, P.-A., {Bingert}, S., \& {Peter}, H. 2013, \aap, 555, A123

\bibitem[{{Chen} {et~al.}(2014){Chen}, {Peter}, {Bingert}, \&
  {Cheung}}]{CPBC14}
{Chen}, F., {Peter}, H., {Bingert}, S., \& {Cheung}, M.~C.~M. 2014, \aap, 564,
  A12

\bibitem[{{Chen} {et~al.}(2015){Chen}, {Peter}, {Bingert}, \&
  {Cheung}}]{CPBC15}
{Chen}, F., {Peter}, H., {Bingert}, S., \& {Cheung}, M.~C.~M. 2015, Nat Phys,
  11, 492

\bibitem[{{Chen, F.} \& {Peter, H.}(2015)}]{CP15}
{Chen, F.} \& {Peter, H.} 2015, \aap, 581, A137

\bibitem[{{Cheung} \& {DeRosa}(2012)}]{CD12}
{Cheung}, M.~C.~M. \& {DeRosa}, M.~L. 2012, \apj, 757, 147

\bibitem[{{Cheung} {et~al.}(2010){Cheung}, {Rempel}, {Title}, \&
  {Sch{\"u}ssler}}]{Cheung_etal10}
{Cheung}, M.~C.~M., {Rempel}, M., {Title}, A.~M., \& {Sch{\"u}ssler}, M. 2010,
  \apj, 720, 233

\bibitem[{{Galsgaard} \& {Nordlund}(1996)}]{GN96}
{Galsgaard}, K. \& {Nordlund}, {\AA}. 1996, \jgr, 101, 13445

\bibitem[{{Gudiksen} \& {Nordlund}(2002)}]{GN02}
{Gudiksen}, B.~V. \& {Nordlund}, {\AA}. 2002, \apjl, 572, L113

\bibitem[{{Gudiksen} \& {Nordlund}(2005)}]{GN05a}
{Gudiksen}, B.~V. \& {Nordlund}, {\AA}. 2005, \apj, 618, 1020

\bibitem[{{Hansteen} {et~al.}(2010){Hansteen}, {Hara}, {De Pontieu}, \&
  {Carlsson}}]{HHDC10}
{Hansteen}, V.~H., {Hara}, H., {De Pontieu}, B., \& {Carlsson}, M. 2010, \apj,
  718, 1070

\bibitem[{{K{\"a}pyl{\"a}} {et~al.}(2016){K{\"a}pyl{\"a}}, {Brandenburg},
  {Kleeorin}, {K{\"a}pyl{\"a}}, \& {Rogachevskii}}]{KBKKR16}
{K{\"a}pyl{\"a}}, P.~J., {Brandenburg}, A., {Kleeorin}, N., {K{\"a}pyl{\"a}},
  M.~J., \& {Rogachevskii}, I. 2016, \aap, 588, A150

\bibitem[{{Klimchuk}(2006)}]{Klim06}
{Klimchuk}, J.~A. 2006, \solphys, 234, 41

\bibitem[{{Mok} {et~al.}(2005){Mok}, {Miki{\'c}}, {Lionello}, \&
  {Linker}}]{MMLL05}
{Mok}, Y., {Miki{\'c}}, Z., {Lionello}, R., \& {Linker}, J.~A. 2005, \apj, 621,
  1098

\bibitem[{{Mok} {et~al.}(2008){Mok}, {Miki{\'c}}, {Lionello}, \&
  {Linker}}]{MMLL08}
{Mok}, Y., {Miki{\'c}}, Z., {Lionello}, R., \& {Linker}, J.~A. 2008, \apjl,
  679, L161

\bibitem[{{Myers} {et~al.}(2015){Myers}, {Yamada}, {Ji}, {Yoo}, {Fox},
  {Jara-Almonte}, {Savcheva}, \& {Deluca}}]{MYJYFJSD15}
{Myers}, C.~E., {Yamada}, M., {Ji}, H., {et~al.} 2015, \nat, 528, 526

\bibitem[{{Oz} {et~al.}(2011){Oz}, {Myers}, {Yamada}, {Ji}, {Kulsrud}, \&
  {Xie}}]{OMYJKX11}
{Oz}, E., {Myers}, C.~E., {Yamada}, M., {et~al.} 2011, Physics of Plasmas, 18,
  102107

\bibitem[{{Pariat} {et~al.}(2015){Pariat}, {Valori}, {D{\'e}moulin}, \&
  {Dalmasse}}]{PVDD15}
{Pariat}, E., {Valori}, G., {D{\'e}moulin}, P., \& {Dalmasse}, K. 2015, \aap,
  580, A128

\bibitem[{{Parker}(1972)}]{P72}
{Parker}, E.~N. 1972, \apj, 174, 499

\bibitem[{{Peter}(2015)}]{P15}
{Peter}, H. 2015, Philosophical Transactions of the Royal Society of London
  Series A, 373, 20150055

\bibitem[{{Peter} {et~al.}(2004){Peter}, {Gudiksen}, \& {Nordlund}}]{PGN04}
{Peter}, H., {Gudiksen}, B.~V., \& {Nordlund}, {\AA}. 2004, \apjl, 617, L85

\bibitem[{{Peter} {et~al.}(2006){Peter}, {Gudiksen}, \& {Nordlund}}]{PGN06}
{Peter}, H., {Gudiksen}, B.~V., \& {Nordlund}, {\AA}. 2006, \apj, 638, 1086

\bibitem[{{Peter} {et~al.}(2015){Peter}, {Warnecke}, {Chitta}, \&
  {Cameron}}]{PWCC15}
{Peter}, H., {Warnecke}, J., {Chitta}, L.~P., \& {Cameron}, R.~H. 2015, \aap,
  584, A68

\bibitem[{{Rappazzo} {et~al.}(2008){Rappazzo}, {Velli}, {Einaudi}, \&
  {Dahlburg}}]{RVED08}
{Rappazzo}, A.~F., {Velli}, M., {Einaudi}, G., \& {Dahlburg}, R.~B. 2008, \apj,
  677, 1348

\bibitem[{{Rempel} \& {Cheung}(2014)}]{RC14}
{Rempel}, M. \& {Cheung}, M.~C.~M. 2014, \apj, 785, 90

\bibitem[{{Stein} \& {Nordlund}(2012)}]{SN12}
{Stein}, R.~F. \& {Nordlund}, {\AA}. 2012, \apjl, 753, L13

\bibitem[{{T{\"o}r{\"o}k} \& {Kliem}(2005)}]{TK05}
{T{\"o}r{\"o}k}, T. \& {Kliem}, B. 2005, \apjl, 630, L97

\bibitem[{{Warnecke} \& {Brandenburg}(2010)}]{WB10}
{Warnecke}, J. \& {Brandenburg}, A. 2010, \aap, 523, A19

\bibitem[{{Warnecke} {et~al.}(2011){Warnecke}, {Brandenburg}, \&
  {Mitra}}]{WBM11}
{Warnecke}, J., {Brandenburg}, A., \& {Mitra}, D. 2011, \aap, 534, A11

\bibitem[{{Warnecke} {et~al.}(2012){Warnecke}, {K{\"a}pyl{\"a}}, {Mantere}, \&
  {Brandenburg}}]{WKMB12}
{Warnecke}, J., {K{\"a}pyl{\"a}}, P.~J., {Mantere}, M.~J., \& {Brandenburg}, A.
  2012, \solphys, 280, 299

\bibitem[{{Warnecke} {et~al.}(2013){Warnecke}, {Losada}, {Brandenburg},
  {Kleeorin}, \& {Rogachevskii}}]{WLBKR13}
{Warnecke}, J., {Losada}, I.~R., {Brandenburg}, A., {Kleeorin}, N., \&
  {Rogachevskii}, I. 2013, \apjl, 777, L37

\bibitem[{{Warnecke} {et~al.}(2016){Warnecke}, {Losada}, {Brandenburg},
  {Kleeorin}, \& {Rogachevskii}}]{WLBKR16}
{Warnecke}, J., {Losada}, I.~R., {Brandenburg}, A., {Kleeorin}, N., \&
  {Rogachevskii}, I. 2016, \aap, 589, A125

\bibitem[{{Wiegelmann}(2008)}]{Wi08}
{Wiegelmann}, T. 2008, \jgr, 113, A12

\end{thebibliography}

\end{document}